\def\comment#1{}

\newcommand{\rp}{\rho_{pp}}
\newcommand{\rn}{\rho_{nn}}
\newcommand{\rd}{\rho_{pn}}
\newcommand{\beg}{\begin{eqnarray}}
\newcommand{\eee}{\end{eqnarray}}

\documentclass[prl,twocolumn,aps]{revtex4-1}
\usepackage{graphicx}
\def\cm#1{}

\begin{document}
\title{Unconventional rotational responses of hadronic superfluids in a neutron star caused by strong entrainment and a $\Sigma^-$ hyperon gap }
\author{ Egor Babaev}
\address{Physics Department, University of Massachusetts, Amherst MA 01003, USA\\
Department of Theoretical Physics, The Royal Institute of Technology 10691 Stockholm, Sweden}
\begin{abstract}
I show that the usual 
model of the rotational response of a neutron star, which predicts rotation-induced neutronic vortices and
no rotation-induced protonic vortices, does not hold (i) beyond a certain threshold of entrainment interaction strength  nor (ii) in case of nonzero
 $\Sigma^-$ hyperon gap. I demonstrate that in both these cases the rotational response involves creation of
phase windings in electrically charged condensate. 
Lattices of {\it bound states of vortices} which are caused these effects
 can (for a range of parameters) strongly reduce the interaction between rotation-induced vortices with magnetic-field carrying superconducting components.
\end{abstract}
\maketitle
\newcommand{\la}{\label}
\newcommand{\aaa}{\frac{2 e}{\hbar c}}
\newcommand{\kA}{{\tilde A}}
\newcommand{\bfx}{{\bf \vec x}}
\newcommand{\bfn}{{\bf \vec n}}
\newcommand{\bfE}{{\bf \vec E}}
\newcommand{\bfB}{{\bf \vec B}}
\newcommand{\bfv}{{\bf \vec v}}
\newcommand{\bfU}{{\bf \vec U}}
\newcommand{\ccc}{{\vec{\sf C}}}
\newcommand{\bfp}{{\bf \vec p}}
\newcommand{\f}{\frac}
\newcommand{\bfA}{{\bf \vec A}}
\newcommand{\non}{\nonumber}
\newcommand{\be}{\begin{equation}}
\newcommand{\ee}{\end{equation}}
\newcommand{\ba}{\begin{eqnarray}}
\newcommand{\ea}{\end{eqnarray}}
\newcommand{\bastar}{\begin{eqnarray*}}
\newcommand{\eastar}{\end{eqnarray*}}
\newcommand{\h}{{1 \over 2}}

Microscopic calculations of  properties of  interior of a neutron star suggest
 a presence of a mixture of superfluid neutrons, superconducting protons and 
 normal electrons 
 (see e.g.  \cite{peth} -\cite{sauls}).
Where protons 
form a type-II superconductor. 
This model suggests that the magnetic field in the star 
should induce a large number of vortices 
in the protonic condensate \cite{peth,ruderman}, while
 rotation should induce neutronic vortex lattice.
Because the magnetic field configuration in a young star is very complicated 
the protonic vortex structure 
is expected to be a complicated vortex tangle \cite{peth,ruderman}.
It is important that in this standard picture for a typical star the average distance between protonic vortices is much larger
than the magnetic field penetration length (which is of order of 10-100fm). Such a vortex tangle 
is indeed a thermodynamically unstable state but estimates
suggest that  a very large time scale is needed to expel these vortices \cite{peth}.

The microscopic calculations also indicate that in a neutron star the effective mass of proton 
is very different from the bare mass \cite{sjoberg}. This implies that there 
is a strong  dissipationless drag effect \cite{andr}: i.e. superfluid velocity of neutronic
condensate drags   superfluid density of protonic condensate and vise versa.
In particular a neutronic circulation  
drags along some density of protons resulting in magnetic field generated by 
neutronic vortex. Therefore a neutronic vortex interacts  magnetically 
 with protonic vortices  \cite{alpar}. 

In the usual model for neutron star, the free energy of the mixture of neutronic and protonic
condnensates has the form:
 \cite{alpar}:
\beg
F=\f{1}{2}\rho_{pp} {\bf v}_p^2+\f{1}{2}\rho_{nn} {\bf v}_n^2+
\rho_{pn} {\bf v}_p \cdot{\bf v}_n +  \f{{\bf B}^2}{2}
\la{f}
\eee
where 
${\bf B}=\nabla \times {\bf A}$ is magnetic field, while 
 ${\bf v}_n=(1/2m_n )\nabla \phi_n$
 and 
${\bf v}_p=(1/2m_p) \nabla \phi_p-(e/m_p) {\bf A}$
are the superfluid velocities of neutron and proton condensates in units $\hbar=c=1$
(generalizations to cases with non-s-wave pairing is straightforward).
Here $m_n\approx m_p=m$ are the bare masses of a neutron and a proton and  $\phi_{n,p}$
are the phases of the corresponding condensates.
The third term in (\ref{f}) represents current-current interaction \cite{andr}.
Because of it the particle
current of one of the condensates (${\bf w}_{p,n}$)  is carried by the superfluid velocity of another:
\beg
{\bf w}_p= \rho_{pp}{\bf v}_p +  \rho_{pn}{\bf v}_n;  \ \ \ {\bf w}_n= \rho_{nn}{\bf v}_n +  \rho_{pn}{\bf v}_p,
\la{gg}
\eee
where $\rho_{pn}=\rho_{np}$ are the drag coefficients which can be expressed via effective masses $m^*_{p,n}$ as follows
 $\rho_{pn}=\rho_{pp}\frac{m^*_p-m_p}{m_p}=\rho_{nn}\frac{m^*_n-m_n}{m_n}$.
 Different microscopic calculations \cite{sjoberg} give for
 $m^*_p$  the
 values ranging $m^*_p\approx 0.3m_p$ to $m^*_p\approx 0.9m_p$.
Thus the drag strength 
can be as high as $\rho_{pn}\approx - 0.7\rho_{pp}$.
From  the eq.(\ref{f}) it follows that the electric current induced by protonic circulation
or by neutronic drag   is given by: 
\be
{\bf J}=e\frac{{\bf w}_p}{m}=
\f{e\rho_{pp} }{m^2}\left(\f{\rho_{pn}}{\rho_{pp} }\nabla \phi_n+
 \nabla \phi_p    -  { e} {\bf A}
\right).
\la{cur}
\ee
When there is a $2\pi$ phase winding in $\phi_p$ (i.e.
for an integral around the vortex core $\oint \nabla \phi_p = 2\pi$), but no phase windings in  $\phi_n$ 
(we denote this vortex as (1,0))
the circulation of electric charge   is  
\be
{\bf J}=
\f{e\rho_{pp} }{m^2}\left( \nabla \phi_p    -  { e} {\bf A}
\right)
\la{curp}
\ee
In the  constant density approximation
the   magnetic field obeys the London equation \cite{pitaevskii}:
${\bf B} + \lambda^2 {\rm curl}  \ {\rm curl} {\bf B}=\Phi_0 \delta ({\bf r})$
where $\Phi_0$ is the magnetic flux quantum
and $\lambda=m/e\sqrt{\rp}$ is the magnetic field penetration length.
Imposing a $2\pi$ winding in $\phi_p$ in eq. (\ref{f})
one obtains a vortex with finite energy  per unit length:
\be
E\approx\int d^2 {\bf r} \left(  \f{1}{2} \f{{\bf J}^2}{e^2\rp} + \f{1}{2} (\nabla \times{\bf A})^2\right)
\approx \left(\f{\Phi_0}{\lambda}\right)^2 \log \f{\lambda}{a}, \nonumber
\ee
where $a$ is the cutoff length associated with vortex core. 

In the case when a $2\pi$ winding is imposed only 
on the neutronic condensates [lets denote it  (0,1)], 
the drag effect produces some protonic current:
\beg
{\bf J}=
\f{e\rho_{pp} }{m^2}\left(\f{\rho_{pn}}{\rho_{pp} }\nabla \phi_n    -  { e} {\bf A}
\right)={e\rho_{pn} }{m^2}\left(\nabla \phi_n    -  { e} \f{\rho_{pp}}{\rho_{pn} }{\bf A}
\right). \nonumber
\eee
Indeed 
this drag current in turn
produces magnetic flux which is  not quantized but is  a linear
function of the drag strength: $\Phi_d = \oint_\sigma  dl {\bf A} = \f{\rd}{\rp}\f{2\pi}{e}$ 
(here the integration is done over the contour $\sigma$ located where  $J\approx 0$).
The main contribution to the energy of (0,1) vortex is associated with the
kinetic energy of the superflow of neutrons.
This contribution is logarithmically divergent as a function of the system size R:
\be
E_{sf}=\int \f{1}{2m^2}{\rn}(\nabla\phi_n)^2 \approx \pi \f{\rn}{m^2}\log\left( \f{R}{a}\right).
\ee
The secondary contribution is associated with the 
drag-induced kinetic energy of the protonic supercurrent 
and the magnetic energy. Because electrically charged
currents are localized at the length scale $\lambda$
(in contrast to neutronic superflow), this contribution is 
much smaller:
\beg
E_{ch}\approx\int d^2 {\bf r} \f{1}{2}\left(   \f{{\bf J}^2}{e^2\rp} +  (\nabla \times{\bf A})^2\right)\approx \left(\f{\Phi_{d}}{\lambda}\right)^2 \log \f{\lambda}{a}.  \nonumber
\eee
If a protonic fluxtube  intersects with a neutronic vortex which carries co-directed flux,
the magnetic energy and kinetic energy of supercurrents rise. 
The energy of the intersection can be estimated as:
\beg
E_{ch}^i\approx \left[ \left(\f{\Phi_{d}+\Phi_0}{\lambda}\right)^2 -\f{\Phi_0^2}{\lambda^2}- \f{\Phi_{d}^2}{\lambda^2} \right] \log \f{\lambda}{a} \approx 
\f{2\Phi_{d}\Phi_0}{\lambda^2}  \log \f{\lambda}{a} \nonumber
\label{interact}
\eee
 According to  \cite{ruderman,link} $E_{ch}^i\approx 5MeV$.
 Because protonic vortex configuration is complicated 
and because protonic vortices are much more numerous than neutronic vortices \cite{peth}, 
there should be strong pinning between the protonic and neutronic vortex matter. 
Much interest in the physics of vortex interaction 
was sparked by the recent calculations in 
 \cite{link} which suggest that the picture of strongly pinned protonic and neutronic vortices may be highly inconsistent
with the slow precession   observed in  a  few isolated pulsars. 
Although most of  pulsars do not have such precession, this inconsistency casted doubts on  the validity of 
the usual neutron star model.
According to \cite{link}, the precession would be possible
only if the  drag strength  was  orders of magnitude
lower than values obtained in microscopic calculations.
Following the work  \cite{link} some alternative scenarios of magnetic response of a neutron star interior were 
proposed.
In \cite{jones},
based on the calculations of $\Sigma^-$ hyperon gap \cite{hyperon},
it was proposed that one has a mixture of two 
oppositely charged condensates which gives further alternatives to type-I/type-II dichotomy \cite{prb05}
(for other aspects of the theory of two-component charged mixture see \cite{naturep,nature,n2}).
The alternative conjecture  put forward was that protons form type-I superconductor \cite{link,zhitn,alford},
suggesting that the magnetic field would penetrate star interior
via  macroscopic normal  domains. 
However it was pointed out in \cite{jones} that the energy difference of a  neutronic vortex located
inside superconducting region versus inside a large type-I domain is quite large, which again leads to 
a strong interaction between
neutronic vortices and flux-carrying domains.
Let me stress that  importantly  the magnetic response of a type-I superconductor is highly nonuniversal.
Generically this magnetic response involves formation of  multi-domains.
The precise shape of the domains depends on a number of factors and is magnetic-history dependent 
\cite{prozorov}.
Importantly a boundary between a normal metal and a superconductor
in magnetic field implies that { there is a magnetic field and
supercurrents within the range of $\lambda$} at this boundary.
Thus even if protons form a type-I superconductor, {  neutronic vortices  will  be strongly interacting with
 the boundaries of normal domains} which typically are plentiful
in a  type-I superconductor   \cite{prozorov}. The strength 
of the resulting pinning of neutronic vortices
will depend on  the domain's structure (which is usually very complicated). 
  Besides that the   mechanism responsible for 
formation of large normal domains in a type-I superconductor is 
assocaited with the dominance of attractive core-core interaction between type-I vortices. However
 this force has the range of the coherence length. For a dilute vortex system with low mobility (as  expected to be the case  in a neutron star) the coalescence 
tendency of type-I vortices (and therefore the tendency to  form large normal domains) would therefore be extremely small. 
So in general, the interaction strength between neutronic vortex lattice and 
flux-carrying structures in type-I case should not
be expected to be dramatically lower than the interaction between neutronic  vortices and
protonic vortices in a type-II case.

Therefore it is an interesting question if there exists a mechanism which, in principle, could lead
to any   significant coupling reduction between neutronic and protonic vortices.
Here I propose  two scenarios which lead to 
such an effect. 
First let me observe that beyond a certain threshold  of  the  drag strength, the usual neutronic vortices are {\it not thermodynamically stable}. That is, the vortex lattices form because 
such a state minimizes the free energy
in a rotating system. The free energy of a (0,1) vortex in a system rotating with the angular velocity ${ \Omega}$ is \cite{pitaevskii}
\beg
F_r^{(0,1)}\approx  \pi \f{\rn}{m^2}\log\left( \f{R}{a}\right) + \left(\f{\Phi_{d}}{\lambda}\right)^2 \log \f{\lambda}{a}
-{ \bf M} { \bf \Omega}, 
\label{g}
\eee
Where  M is the vortex momentum \cite{pitaevskii}:
\be
M= \int d^3{r} [ \frac{r}{m}\rd|({\nabla\phi_n} - e \frac{\rp}{\rd} {\bf A})| + \frac{r}{m} \rn |{\nabla\phi_n}|]
\label{M}
\ee
In eq. (\ref{M}) the first contribution comes from protonic circulation, while the second term corresponds to the neutronic circulation.
The supercurrent of protons is concentrated around the vortex within the range of penetration length  (of order of  $10$ or $100$fm in the standard picture), 
while the characteristic size of the star is of order of 10 km. Therefore 
the protonic contribution $ \int d^3{r}  ({r}/{m}) \rd|{\nabla\phi_n} - e(\rp/\rd){\bf A})|$ 
 in the vortex momentum (the last term in Eq. (\ref{g})) is much smaller than the neutronic contribution (which is not exponentially localized).
Note also that this small contribution  has the role of the {\it energy penalty} because $\rd<0$
and thus protons are circulating in the direction opposite to the neutronic circulation.  
Let us now rewrite the equation (\ref{f}) as
\beg
F&=&\int d{\bf r}
 \f{1}{2}
\biggl[
\left( \rn - \f{\rd^2}{\rp} \right) 
(\nabla \phi_n)^2 + \nonumber \\
&&  \rp\biggl(\nabla\phi_p+\f{\rd}{\rp}\nabla\phi_n -e{\bf A}\biggr)^2 
+{(\nabla \times {\bf A})^2}\biggr]
\label{separated}
\eee
Here we separated  the energy associated with the electrically
neutral superflow (given by the first term). It is associated with the phase gradient  decoupled from ${\bf A}$.
The remaining terms describe the kinetic energy of the electrically charged currents
 (represented by the phase gradients coupled to vector potential ${\bf A}$)  and the magnetic energy.
In a free energy of (0,1) vortex in a rotating frame, there is (i) a free energy penalty coming from the negative-drag-induced momentum of protons,
and (ii) energy penalties from the kinetic energy of protonic currents and magnetic field.
Let me observe that the simplest vortex with phase winding only in the neutronic phase is not thermodynamically stable
 when  $|{\rd}|/{\rp} > 1/2$. Thermodynamic stability is achieved when the free energy is minimized
 with respect to all the degrees of freedom which include  phase windings. 
Previously it was assumed that rotation induces vortices
with only neutronic phase winding \cite{alpar}. However from the
eq. (\ref{separated}) we can see that   if there is a  $2\pi$ winding in $\phi_n$, 
and  $-{\rd}/{\rp} > 1/2$, the system can  minimize the
free energy 
 in a rotating frame by creating an additional  $2\pi$ winding in the protonic phase $\phi_p$ (around the same core).
Lets denote such composite vortices (1,1). Its
 free energy is
\be
F_r^{(1,1)} \approx \pi \f{\rn}{m^2}\log\left( \f{R}{a}\right) + \left(\f{\Phi_0 (1+\f{\rd}{\rp})}{\lambda}\right)^2 \log \f{\lambda}{a}
-{ M} { \Omega}
\label{g2}
\ee
Note that microscopic calculations \cite{sjoberg} estimate  the drag strength in a neutron star
can be as high as ${\rd}/{\rp} \approx -0.7$.
Let me observe that with {\it increased} drag strength beyond the threshold $-{\rd}/{\rp} >1/2$,
the magnetic energy and the energy of protonic currents around  a rotation-induced (1,1) vortex {\it decreases}. 
Moreover the additional phase winding does not  affect the dominant neutronic
negative contribution  (coming from  $-  M\Omega$) to the 
free energy   (\ref{g}) . 
Also because the protonic current of (1,1) vortex  
\be
{\bf J}^{(1,1)} \propto 
  \left( \left[1-\f{|\rd|}{\rp}\right]\nabla\phi_n -e{\bf A}\right)
 \ee
has a circulation opposite to that  of a (0,1)  vortex   
\be
{\bf J}^{(0,1)}  \propto \left(-\f{|\rd|}{\rp}\nabla\phi_n -e{\bf A}\right)
\ee
 it carries an opposite  momentum.
Thus  instead of a positive free energy penalty 
 $-  {\bf M}\cdot{\bf \Omega}$ it produces a negative free energy gain in this term.
{\it  Therefore $F_r^{(1,1)}<F_r^{(0,1)}$ for  $-{\rd}/{\rp} > 1/2$
and thus the (0,1) neutronic vortices are not thermodynamically stable in this regime.
Instead a superfluid supports rotation by creating a lattice of composite vortices (1,1). 
}
Observe that in general, for a given superfluid momentum,
 there is also a possibility to minimize the energy of charged currents
for smaller ratios of $|{\rd}|/{\rp}$ by creating a $2\pi N$ winding in $\phi_n$ 
and compensating the contribution of   $\nabla\phi_n$  in the second term in (\ref{separated})
by a $2\pi M$ ``counter-winding" in $\phi_p$. But in the case of a vortex with a multiple winding in $\phi_n$ the first term 
in (\ref{separated}) depends quadratically on $N$ while the last term depends on it
linearly. Because in a neutron star, the $\rho_{nn}$ is very large compared to  $\rho_{pn}$
and $\rho_{pp}$  the
composite vortices with multiple windings in $\phi_n$, like e.g. (2,1) should not be thermodynamically stable.

Another physical consequence of the possible formation of
composite (1,1) vortices is that for $-{\rd}/{\rp} >1/2$ with {\it increased} drag strength $|{\rd}|$
 there is a {\it decrease} in the protonic current and magnetic energy contributions to the
energy cost of an intersection between the  rotation induced vortex and protonic (1,0) flux tube. 
For nearly parallel vortices it is given by
\be
E_{ch}\approx  {{2\Phi_0^2(1+{\rd}/{\rp})}}{\lambda^{-2}}  \log ({\lambda}/{a}). 
\label{interact2}
\ee
Also the (1,1) vortices should have reduced interaction with the 
boundaries of the normal domains (if protonic condensate is type-I).
To estimate more accurately vortex-vortex interaction potential, it is important to observe 
that the composite (1,1) vortex has a core in protonic condensate due to protonic  phase winding.
This gives rise to the attractive core-core interaction with the range of coherence length \cite{Kramer}
(which originates
from minimization of the energy cost of suppression of the protonic order parameters in the cores).
Therefore if protons form type-II condensate then  for $\rd/\rp>0.5$ the interaction between approximately co-directed (1,1) and (1,0) vortices is further minimized by
core-core interaction because both (1,1) and (1,0) vortices have zeros of protonic condensate.

There is another scenario leading to appearance of composite vortices and possible electromagnetic coupling reduction in a neutron star.
It has been discussed that a star can have another electrically charged
condensate associated with $\Sigma^-$ Cooper pairs \cite{jones,hyperon}. 
Let me observe that from the theory of two-component charged mixture \cite{naturep}
it follows that  the presence of the second charged component alters qualitatively the 
rotational response of the system (if intercomponent Josephson coupling is negligible \cite{naturep}).
 That is, in the free energy functional of such a system, one can   form a gauge-invariant combination
of the phase gradients $\propto (\nabla (\phi_p + \phi_\Sigma))$ which is decoupled from the vector 
potential  \cite{nature,naturep}  (here $\phi_\Sigma$ is the phase of hyperon condensate).
It means that if another charged component is present, then there is an additional superfluid mode (i.e. besides the neutronic) associated with 
co-flow of oppositely charged neutronic and hyperonic Cooper pairs. Such co-flows transfer mass without charge transfer.
 From the calculations in \cite{naturep,nature} it follows
the system has  two kinds of  vortex excitations which carry
identical   momenta associated with these co-flows. These vortices have phase windings $(\Delta \theta_p = 2\pi, \Delta \theta_\Sigma=0)$
and $(\Delta \theta_p =0, \Delta \theta_\Sigma=2\pi)$. On the other hand these vortices
carry different fractions of magnetic flux quanta:
\be
\Phi_i=\pm\frac{|\Psi_i|^2}{m_i}\left[ \frac{|\Psi_p|^2}{m_p} +\frac{|\Psi_\Sigma|^2}{m_\Sigma}\right]^{-1} \Phi_0,
\ee
where $|\Psi_{(p, \Sigma)}|^2$ and $m_{(p, \Sigma)}$ are protonic and hyperonic condensate's number densities and masses.
The density $\nu$ of rotation-induced vortices  in a mixture of charged condensates depends on the angular velocty $\Omega$ 
as \cite{naturep}
\be
\nu=\frac{(m_p+m_\Sigma)\Omega}{\pi\hbar}.
\ee 
Therefore these vortices should be more numerous than  neutronic vortices (observe that this effect takes
place no matter  how small  the superfluid density of the 
 second charged component is). At the same time the phase
stiffness of the composite superfluid mode  is tiny $\frac{|\Psi_p|^2}{m_p}\frac{|\Psi_\Sigma|^2}{m_\Sigma}\left[ \frac{|\Psi_p|^2}{m_p} +\frac{|\Psi_\Sigma|^2}{m_\Sigma}\right]^{-1}$ \cite{nature,naturep}. It is  indeed much smaller than the phase stiffness of the neutronic condensate
and therefore ordering energies of the rotation-induced protonic/hyperonic vortex lattice are much smaller than those of 
neutronic vortex lattice. In this system therefore there will be a competition between minimization of vortex lattice 
ordering energy versus minimization of the kinetic energy of charged currents and magnetic energy.
Both the neutronic vortices and  the rotation-induced vortices in  a charged condensate mixture \cite{nature} carry fractional magnetic flux. It is known from the  other examples \cite{dahl,naturep} that this kind of competitions
results in formation of composite vortex lattices  if the ordering energy of one of the vortex sublattices is low (it is 
indeed the case in the system in question). 
 In particular one neutronic vortex can anchor one or several rotation-induced
vortices  in the charged condensates mixture. There is certainly a range of parameters where the resulting composite vortices
carry very small  magnetic flux. This would result in a different scenario of electromagnetic coupling  reduction
of a rotation-induced vortex lattice and magnetic field carrying structures \cite{mf}.  Observe also that in a mixture the London law is modified \cite{naturep}.

In conclusion, in the usually assumed picture, a  
rotation of a neutronal star induces neutronic vortex lattices,
and magnetic field produces
protonic fluxtubes.  It is  also assumed in the usual picture 
that no vortices in the protonic condensate
can be induced by rotation since single-component charged condensate reacts to rotation via the London law.
In this work it was shown that a mixture of hadronic superfluids in a neutron star has a qualitatively different
rotational  response  if drag strength exceeds a certain threshold  $\rd/\rp<-1/2$
(microscopic calculations suggest that $\rd/\rp$ can be as large as $-0.7$)
or if several charged condensates are present. In both these cases  rotation
actually does produce phase windings in a charged condensate 
i.e. protonic (or $\Sigma^-$-hyperonic) vortices.
Both these scenarios,  may lead to a strong reduction 
of electromagnetic coupling between a rotation-induced vortex
lattice and magnetic-flux carrying structures.
If the resulting interaction strength falls below the temperature scale,
in some region of a neutron star, it can lead to decoupling of superfluid and flux-bearing components.


\end{document}